%% file: HAIM-technical.tex
\documentclass[lettersize,journal]{IEEEtran}

\usepackage{hyperref}
\usepackage{amsmath}
\usepackage{authblk}
\usepackage{geometry}
\usepackage{lipsum}
\usepackage[font=footnotesize,labelformat=simple]{subcaption}
\usepackage{cleveref}

\usepackage{algorithm}
\usepackage[noend]{algorithmic}

\usepackage{soul}

\usepackage{listings}

\input{header}

\begin{document}
\thispagestyle{fancy}

\title{Technical Report: Hybrid Autonomous Intersection Management}

\author[1]{Aaron Parks-Young}
\author[1]{Guni Sharon}
\affil[1]{Computer Science and Engineering, Texas A\&M University, 3112 TAMU, College Station, TX 77843-3112, United States}
\renewcommand\Authands{ and }
\date{\today}

\maketitle

This document provides technical details regarding the Hybrid-AIM simulator that was used in~\cite{sharon2017protocol} and in~\cite{HAIM_Parks-Young_Sharon}.

\section{Survey of Relevant Publications} \label{other_related_work}
Since Dresner and Stone’s 2008 work, the research field of autonomous vehicle coordination and autonomous intersection management has been very active. However, while many of the methodologies in the specific area of autonomous intersection management show promise, they typically fail to consider human vehicles, show results with only simplistic simulations with limited relevancy to the real-world, or fail to leverage currently deployed traffic management methods. Thus, it is difficult to imagine fielding such approaches without significant adjustments. This section provides an overview of some of the vast amount of work that has been done. The reader is referred to \cite{namazi2019intelligent} and \cite{chen2015cooperative} for a full survey regarding reservation-based intersection management systems. 

A number of works approach autonomous intersections from a purely control and/or optimization point of view~\cite{riegger2016centralized,kim2014mpc,malikopoulos2020optimal, murgovski2015convex}. These approaches, however, are limited to connected vehicles with high control precision. This constraint prevents them from being applied to situations with human operated vehicles. Some work has begun to emerge which reformulates the optimization approach to account for human operated vehicles, such as in \cite{liu2018safe}. However, Liu et al. provide no empirical analysis and do not explicitly consider modern signal control methods such as signal actuation.

Another potential improvement to intersection management schemes are approaches known as platooning and batch reservations~\cite{au2011enforcing,bashiri2018paim, jin2013platoon}. While batch reservations and platooning support may be helpful as components of a reservation protocol, the major advantages of these methods will not be observable in implementation until significant CAV penetration levels are realized. Some game theoretic/incentive based~\citep{sayin2018information, wei2018intersection} or auction based~\citep{vasirani2012market, carlino2013auction, levin2015intersection} techniques for allocating reservations have also been proposed. These methods rely on connected vehicles' abilities to communicate and follow granted reservations as well as communicating personal preferences, which means this class of approach is not suitable for vehicles with no communication capabilities. Some areas of work forgo centralized intersection management architecture entirely and focus on employing distributed methods~\cite{vanmiddlesworth2008replacing,mladenovic2013self}. However, these works are only demonstrated to work with lower traffic volumes. Additionally, distributed approaches face significant safety and applicability problems when HVs are considered. Another class of work suggested incorporating reinforcement learning as part of intersection management systems~\cite{mirzaei2017fine,wu2019dcl}. Such protocols rely on less practical sensing assumptions such as a fully observable environment in terms of movement intention of vehicles, vehicle speed, vehicle position, and per lane queue lengths.

\begin{figure}[]
	\centering
	\includegraphics[width=3in]{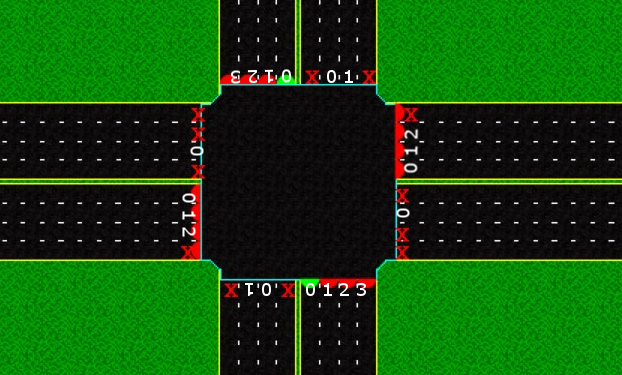}
	\caption{Example intersection in the modified AIM Simulator including removed lanes from the original, symmetric intersection layout (marked by red `x') and relative lane indices superimposed.}
	\label{fig:laneclosureexample}
\end{figure} 

\begin{figure*}[]
\centering
\begin{center}
\includegraphics[width=15 cm]{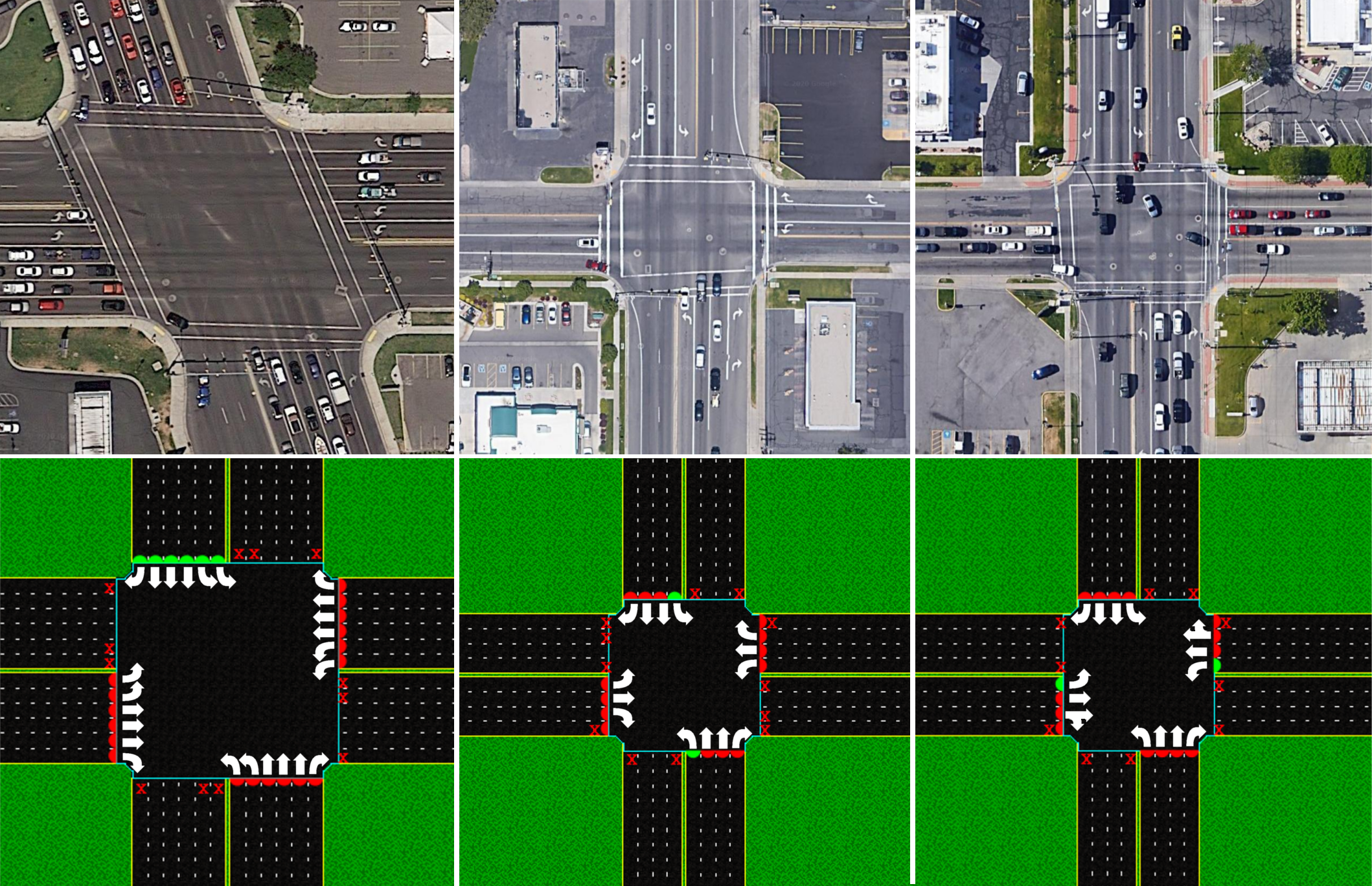}
\end{center}
\caption[Intersection \#6303, \#7204, and \#7381 above the respective AIM Simulator representations.]{Intersection \#6303 (top left, State St. at 800 North), \#7204 (top center, 900 East at 5600 South), and \#7381 (top right, 5600 West at 3500 South) above the respective AIM Simulator representations. Current/real turning assignments are superimposed on top of the AIM Simulator representations. Satellite Imagery: Imagery  \copyright~2020 Maxar Technologies, State of Utah, Map data \copyright~2020. Courtesy of Google~\citep{maps}.}
\label{fig:inters}
\end{figure*} 

\subsection{Simulation Environment}

\subsubsection{Demand, intersection layout, and turning profiles} \label{demand_lanes_and_turning_in_aim}

Our modified AIM Simulator uses a comma delimited file generated by UDoT's ATSPM system~\citep{udotatspm} to schedule vehicle arrivals for a given intersection. Vehicles' arrival times are randomly sampled using a uniform distribution 
within ``buckets'' of time (e.g., 5 minutes) specified by each row within the comma delimited file. Specification via file allows for finely controllable arrival rates in simulation and makes the simulator capable of simulating peaks and valleys in traffic demand over a period of time. The ability to generate demand based on real historical data or even based on finely detailed artificial data makes the simulator capable of modeling more realistic scenarios as compared to earlier versions.

In previous versions, the AIM Simulator assumed a symmetric intersection layout. That is, all (four) directions of travel have the same number of incoming and outgoing lanes and all roads have the same speed limits. In order to model single intersections that are not totally symmetric in this manner, the simulator is now modified to accommodate a prespecified number of inbound and outbound lanes per road as well as some restrictions regarding what trajectories are considered valid within the intersection (incoming to outgoing lanes). 
These specifications as well as the assignment of speed limits are done in a configuration file. An example of an asymmetric intersection in the adapted AIM Simulator is depicted in Figure~\ref{fig:laneclosureexample}. The adapted intersection uses a symmetric intersection as a basis and then removes excess lanes (marked by a red `x' in the figure).

Another consideration in modeling realistic scenarios is the actions vehicles may take when entering an intersection. Modern intersections specify allowed turning movements (e.g., left, right, straight) for vehicles in each lane. The modified AIM Simulator, constructed for this article, allows assignment of such turning profiles to different types of vehicles (CAVs vs. HVs). Note that, turning profiles only specify the general actions a vehicle may take when proceeding through an intersection and they should not be confused with conditions dictating whether a vehicle may enter the intersection at all, such as a requirement for a green light (for HVs) or an approved reservation (for CAVs). The specification of turning profiles for vehicles in the modified AIM simulator are made in the same configuration file that determines the number of lanes and the speed limits for roads serviced by the simulated intersection.

\subsubsection{Signal timing} \label{signal_timing_in_aim}
\label{signals_for_humans}
The AIM Simulator's ability to model traffic signal progression was also extended for this article. Rather than a fixed signal timing scheme, the modified AIM Simulator allows for a ring and barrier style signal phase specification via an appropriate configuration file. Additionally, a flag may be set on simulation start to enable signal actuation using the specified progression and timing.
Users can easily specify (as shown in Configuration File Excerpt \ref{signaltimingconfig}) the \textbf{gap extension} (green time extension following a vehicle detection or `actuation'), maximum phase intervals, and minimum phase intervals. Users may choose to enable actuation detection, to use an adaptive signal timing scheme if desired, or use both.
Tables 5-6 and 5-10 from the U.S. Department of Transportation Federal Highway Administration's 2008 Traffic Signal Timing Manual~\citep{trafficsignaltimingmanualchap5} are used in this article when adaptive timing is enabled in order to determine the relevant values at a given time. These values are hard coded into the simulator (accessible in the ``GapExtensionTable'' and ``MaximumGreenTable'' Java files in the modified simulator's code base at \url{https://github.com/Pi-Star-Lab/Improved-H-AIM}). 

\subsubsection{Benchmark intersections}
Using readily available data from UDoT's ATSPM system~\citep{udotatspm}, three real intersections in Utah were selected and modeled for use in each simulation: State St. at 800 North (\#6303), 900 East at 5600 South (\#7204), and 5600 West at 3500 South (\#7381). 
These intersections are modeled based on the number of incoming lanes, the number of outgoing lanes, and the speed limits for each road. Signal progression is approximated as a somewhat standard 8 phase model (though signal timing and phase order/pairings vary). Incoming and outgoing lane configurations were gathered using Google Maps~\citep{maps}. Speed limits were gathered through a combination of Google Maps and information made available through Utah DoT~\citep{udotopendata}. Overhead photos of the intersections may be seen in Figure~\ref{fig:inters} above a screenshot of their respective AIM Simulator representations. Note that the simulator models all intersections as a symmetric ($90^{\circ}$) cross. 
We assume the simulated traffic patterns are flexible enough to handle relatively small discrepancies in the intersections' layouts, but intersections of more irregular
shapes may need to be modeled explicitly.  Extending the simulator in
that direction is an important direction for future work, especially if the
H-AIM protocol is to be applicable to various types of intersections.

\begin{figure}[]
\centering
\begin{center}
  \includegraphics*[width=7.7 cm]{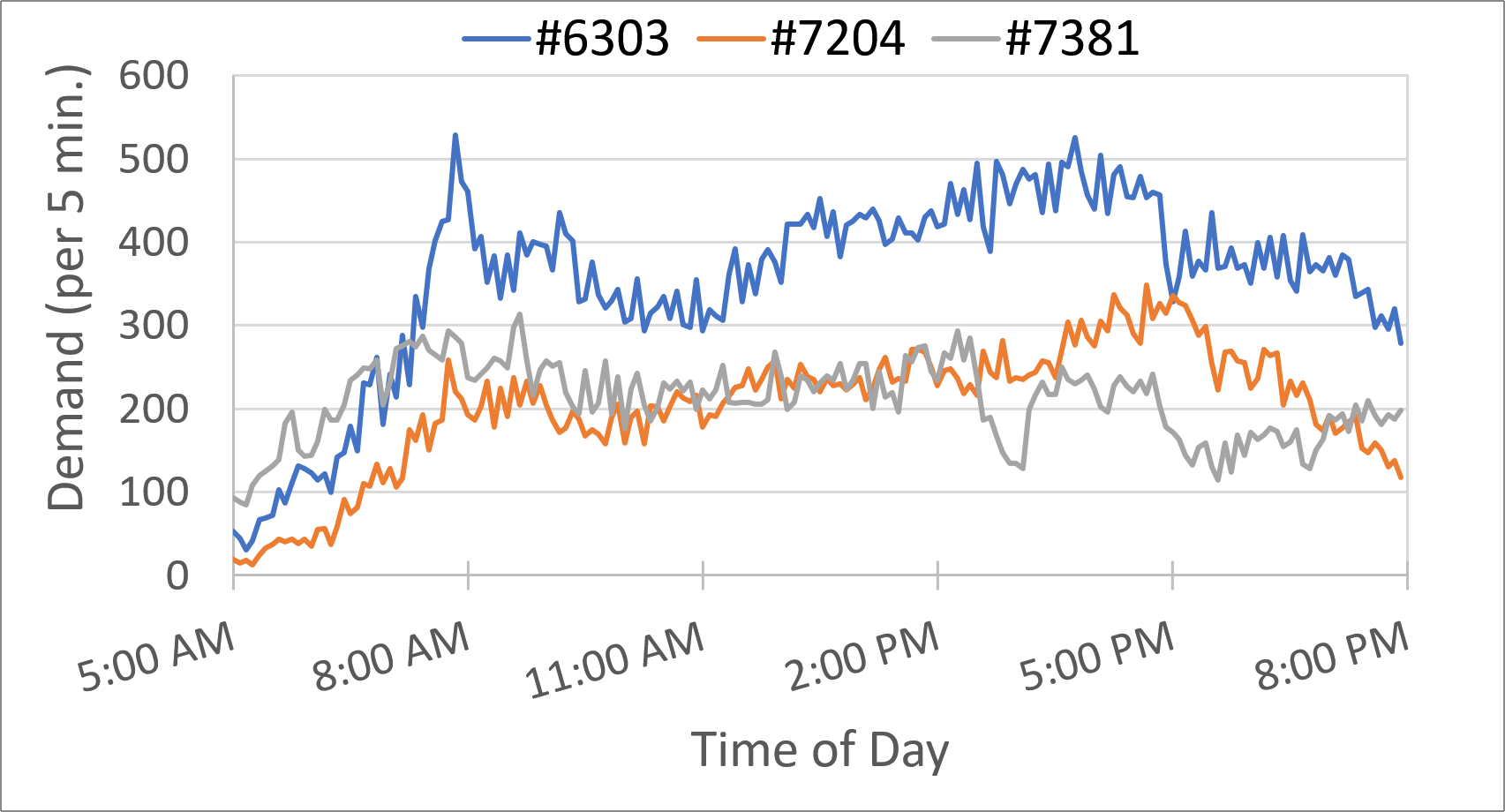}
\end{center}
\caption[Demand in terms of vehicles plotted against time for intersections \#6303, \#7204, and \#7381.]{Demand in terms of vehicles (aggregated in 5-minute chunks) plotted against time for intersections \#6303 (plot on top in blue, starting in the middle of the three), \#7204 (slight upward trend in orange, starting at the bottom of the three), and \#7381 (fairly steady trend in gray, starting at the top of the three).}
\label{fig:demand}
\end{figure} 

Traffic demand data for simulations is taken from UDoT’s ATSPM system for
Friday, March 8, 2019. This date was selected because the affiliated demand demonstrates different traffic patterns for the intersections, which allows for more insightful comparison as opposed to other days where the patterns for intersections are more similar.
Traffic was considered between 5:00 AM inclusive and 8:00 PM exclusive. The specific time window was chosen with the intent of including peak traffic hours for both the morning and the evening. Figure \ref{fig:demand} shows the demand on March 8, 2019 for all three intersections smoothed across 5-minute intervals. Table~\ref{tab:vlh} presents the average, peak (in a sliding full hour window), and low (in a sliding full hour window) vehicles per hour per lane (VLH) for the same three intersections.

\begin{figure*}[]
  \begin{minipage}{\textwidth}
	\begin{lstlisting}[language=Xml, caption=Signal Timing Specification, label={signaltimingconfig}]
<?xml version=``1.0'' encoding=``UTF-8''?>
<root>
        <!--Road direction outbound, -->
        <!--cross (``c'') turn (left in US) turn and/or through-->
        <!--(``t''), gap timeout, min time, max time-->
        <green>N, c, 5, 4, 6.339671678</green>
        <yellow>N, c, 4</yellow>
        <red>N, c, 3</red>
        <green>S, t, 5, 4, 59.11608376</green>
        <barrier id=``b1''></barrier>
        ...
    </ring>
    <ring>
    ...
    </ring>
    <barrier id=``b1''>4, 3</barrier>
    <barrier id=``b2''>4, 3</barrier>
</root>
\end{lstlisting}
  \end{minipage}
\end{figure*}

\subsubsection{Configuration Files}

Configuration File Excerpt~\ref{signaltimingconfig} shows how signal timing is specified in a ring and barrier fashion for the improved AIM Simulator. This configuration file is a standard XML file using $<$green$>$, $<$yellow$>$, and $<$red$>$ tags to represent the various timings of phases within a pair of $<$ring$>$ tags. The document may have multiple $<$ring$>$ tag pairs specified. The $<$barrier$>$ tag is both used to define a barrier and place it within a ring, depending on the position of the tag within the file. All tags should be contained within the $<$root$>$ tags and the XML encoding and specification tag should be at the front of the document to enable correct parsing by the simulator.

\begin{table}[]													
	\begin{center}
		\begin{tabular}{|c|r|r|r|r|r|r|}
	\hline & \textbf{\#6303}&\textbf{\#7204}&\textbf{\#7381}\\
\hline \textbf{VLH Average}&177.33&174.34&179.87\\
\hline \textbf{VLH Peak}&239.04&273.21&234.21 \\
\hline \textbf{VLH Low}&39.08 & 26.29& 111.43\\
\hline \textbf{Lanes}&24 &14 &14\\
\hline
 \end{tabular}
 \vspace{2mm}
\caption{The average, peak (in a sliding full hour window), and low (in a sliding full hour window) vehicles per hour per lane (VLH) for each intersection along with the number of lanes entering the intersection.}
 \label{tab:vlh}
 \end{center}
\end{table}

Within each pair of $<$green$>$ tags, there is first a series of 2 letters. The first letter represents one of four cardinal directions of travel out of \{`N': North, `E': East, `S': South, `W': West\} for the road for which the signal applies. The second value represents the turning movement for which the signal should apply. Currently, the simulator supports any of the following phases \{`c': cross heading traffic (left turn in the U.S), `t': through and non-crossing turn (right in the U.S.), `ct': cross and through and non-crossing turn\}. The third, fourth, and fifth values represent the gap timeout, min green duration, and max green duration respectively for actuated control. The two next lines with tags $<$yellow$>$ and $<$red$>$ should share the same direction and allowed turning movement parameters as the leading $<$green$>$ tag they follow. The $<$yellow$>$ and $<$red$>$ tag pairs both function similarly to the $<$green$>$ tag pairs, but require that only the duration be specified for the respectively yellow or red signal. Note that the simulator functions on discrete time steps of 0.02 seconds by default, and so updates to signals only occur every 0.02 seconds. Each $<$ring$>$ tag begins with a green signal and progress through yellow and red before moving to the next phase within the ring.

 Barriers are identified by the $<$barrier$>$ tag with a specific id attribute. When included within a ring, on top of enforcing phase progression constraints as the name ``barrier'' would suggest, this tag functions as both a $<$yellow$>$ and $<$red$>$ tag with timing based off of the specific barrier's definition. A barrier may be defined by including the $<$barrier$>$ tag at the same level in the document as the $<$ring$>$ tags. Each barrier (differentiated by the id attribute in the tag) should be referenced in each ring at appropriate points in the phase progression (File Excerpt~\ref{signaltimingconfig} for an example).

\begin{figure*}[]
  \begin{minipage}{\textwidth}
  
\begin{lstlisting}[language=Clean, caption=Traffic Arrival Data,  label={trafficspec}]
EAST, WEST, NORTH, SOUTH
L,T,R,Total,L,T,R,Total,L,T,R,Total,L,T,R,Total,Vehicle Total
5:00 AM,1,1,1,3,0,2,0,2,1,3,1,5,1,7,1,9,19
5:05 AM,1,1,1,3,1,3,1,5,0,4,1,5,0,2,0,2,15
5:10 AM,0,2,1,3,2,2,1,5,0,7,0,7,0,3,0,3,18
5:15 AM,2,1,1,4,0,1,0,1,0,5,1,6,0,2,0,2,13
5:20 AM,1,0,1,2,2,0,2,4,3,8,2,13,0,5,0,5,24
5:25 AM,2,0,1,3,2,2,2,6,1,17,1,19,1,4,0,5,33
...
7:30 PM,13,13,0,26,6,12,3,21,7,27,4,38,1,56,5,62,147
7:35 PM,11,7,4,22,6,19,3,28,8,34,8,50,4,45,10,59,159
7:40 PM,5,13,9,27,3,14,3,20,4,31,6,41,6,50,7,63,151
7:45 PM,7,11,7,25,3,10,1,14,3,31,8,42,4,43,2,49,130
7:50 PM,10,13,5,28,5,9,7,21,6,30,1,37,2,46,4,52,138
7:55 PM,1,9,3,13,8,4,2,14,2,33,4,39,2,43,7,52,118
\end{lstlisting}
  \end{minipage}
\end{figure*}

Configuration File Excerpt~\ref{trafficspec} shows how the improved AIM Simulator ingests information about traffic arrivals in a comma delimited format (CSV). This type of data format is produced by the Utah Department of Transportation’s Automated Signal Performance Measures (ATSPM) system. The first row dictates the order of directions of travel which spawned vehicles are assigned. The next row dictates the order of turning actions to be parsed for each road based on the first row. Thus, in Configuration File Excerpt~\ref{trafficspec}, the first series of ``L,T,R'' represents the left, through, and right turning directions for vehicles which spawn heading east in the simulator. The next series of ``L,T,R'' is for westbound vehicles, and so on. Note that if a compound action is encountered, such as ``TR'', the simulator will randomly split vehicles spawning amongst the composing turning actions using a uniform distribution. ``Total'' acts as a delimiter between travel directions on the second line and the associated value in each line of data is ignored. The ``Vehicle Total'' values are ignored as well. The rest of the file is comprised of multiple rows of arrival data.

Each row of arrival data consists of a timestamp followed by a series of numbers. The difference between timestamps on each line is used by the simulator to determine the span of time in which it may spawn the vehicles described for each line. The simulator requires that the difference in time between each line of data is the same and that at least 2 rows of data exist so that the time span may be determined automatically. Note that it is not currently recommended to use data which spans across multiple days, as testing of the simulator's parsing capability in this situation has not been performed. The numbers following the timestamp on each line correspond to the series of turn movements for each road on the second line of the file. Thus, the sequence ``1,1,1'' on the first line of data represents that 3 vehicles heading east should spawn in the first 5 simulation minutes with one choosing to turn left, one choosing to proceed straight, and one choosing to turn right. No spaces should occur on data lines, except for within the timestamp as shown.

\begin{figure*}[]
  \begin{minipage}{\textwidth}
  
\begin{lstlisting}[language=Xml, caption=Intersection Model Specification, label={intersectionspec}]
<?xml version=``1.0'' encoding=``UTF-8''?>
<intersection>
    <!--Direction of travel, incoming, outgoing, speed in m/s,-->
    <!--maximum future time for reservation in seconds-->
    <road>EAST, 3, 1, 13.4, 14.925373134328358208955223880597</road>
    <road>SOUTH, 4, 2, 20.1, 9.950248756218905472636815920398</road>
    <road>WEST, 3, 1, 15.6, 12.820512820512820512820512820513</road>
    <road>NORTH, 4, 2, 20.1, 9.950248756218905472636815920398</road>
	
    <!--Connections-->
    <direction>
        <from_to>EAST, EAST</from_to>
        <vehicle type=``HUMAN''>(1,0)</vehicle>
        <vehicle type=``AUTO''>(1,0)</vehicle>
    </direction>
    <direction>
        <from_to>EAST, NORTH</from_to>
        <vehicle type=``HUMAN''>(0,0)</vehicle>
        <vehicle type=``AUTO''>(0,0), (1, 1)</vehicle>
    </direction>
    <direction>
        <from_to>EAST, SOUTH</from_to>
        <vehicle type=``HUMAN''>(2,1)</vehicle>
        <vehicle type=``AUTO''>(1,0), (2,1)</vehicle>
    </direction>
    ...
    <direction>
        <from_to>SOUTH, SOUTH</from_to>
        <vehicle type=``HUMAN''>(1, 0), (2, 1)</vehicle>
        <vehicle type=``AUTO''>(1, 0), (2, 1)</vehicle>
    </direction>
    <direction>
        <from_to>SOUTH, WEST</from_to>
        <vehicle type=``HUMAN''>(3, 0)</vehicle>
        <vehicle type=``AUTO''>(3, 0)</vehicle>
    </direction>
    <direction>
        <from_to>SOUTH, EAST</from_to>
        <vehicle type=``HUMAN''>(0,0)</vehicle>
        <vehicle type=``AUTO''>(0,0)</vehicle>
    </direction>
</intersection>
\end{lstlisting}

  \end{minipage}
\end{figure*}

Configuration File Excerpt~\ref{intersectionspec} shows an example file defining the allowed turning actions per lane, the lanes entering per road, the lanes exiting per road, speed limits per road in meters per second, and the maximum future time allowed for reservations per road in seconds. The file is an XML file, and so the XML encoding and specification tag should be at the front of the document to enable correct parsing by the simulator. The $<$intersection$>$ tag pair following that tag is the root level tag. Within the $<$intersection$>$ tag pair are $<$road$>$ tags and the $<$direction$>$ tags along with their sub-tags.

Each $<$road$>$ tag contains 5 items. First is the direction of travel for a road. The directions chosen should be the same cardinal directions as in Configuration File Excerpt~\ref{trafficspec}. Second is the number of incoming lanes for the direction of travel followed by the number of outgoing lanes. Fourth is the speed in meters per second. Lastly is the maximum future time in seconds for which a reservation may be acquired for vehicles arriving on that road.

Next is a series of $<$direction$>$ tag pairs and the sub-tags contained within. Each pair of $<$direction$>$ tags defines a turning action and from which lanes the turning action may be taken for a particular road. The first sub-tag is the $<$from\_to$>$ tag which dictates the cardinal direction from which a vehicle is traveling and to which the vehicle will be departing in order for the simulator to determine what turning action is being taken (left, right, through). The `from' direction comes first and then is followed by the `departure' direction. These directions are split by a comma. The next sub-tags are the $<$vehicle$>$ sub-tags which contain a ``type'' parameter. The $<$vehicle$>$ tags define which lanes for each vehicle type may be used to perform the turning action defined by the $<$from\_to$>$ tag. The type parameter dictates the vehicle type and the pairs of numbers contained within the $<$vehicle$>$ tag pair define the mapping of lanes. The lane mapping defines which incoming lanes a vehicle may use to enter the intersection before departing on particular outgoing lanes. This is currently a 1-to-1 mapping (per incoming road) with the first number being the relative index of the lane from the left on the incoming road (so, 0 would be the leftmost lane, 1 is the next lane to the right, etc.) and the second number being the relative index from the left on the outgoing road. 

Putting these all together, it can be observed that the first $<$direction$>$ tag in Configuration File Excerpt~\ref{intersectionspec} defines that all autonomous and human vehicles heading straight on the eastbound road must use the second lane from the left to proceed straight onto the leftmost lane on the outbound side of the intersection. 
Figure~\ref{fig:laneclosureexample} depicts the lane indices for an example intersection and might allow better intuition of the meaning of Configuration File Excerpt~\ref{intersectionspec}.

\begin{table*}[]													
	\begin{center}
		\begin{tabular}{|c|r|r|r|r|r|r|r|r|r|}
	\hline		&	\multicolumn{2}{|c|}{\textbf{\#6303}}&\multicolumn{2}{|c|}{\textbf{\#7204}}&\multicolumn{5}{|c|}{\textbf{\#7381}}\\
	\cline{2-10}CAV Ratio & \multicolumn{9}{|c|}{Turning Policy Variation} \\
	\cline{2-10}	 &C, C&P, C&C, C&P, C&C, C&P, C&C, R&P, R&R, R\\
	\hline 0 &42.6&42.7 &*83.1 &*83.1 &*95.6 &*95.2 & *132.1&*131.7 &*132.5 \\
	 0.01 & 42.4&42.5 & *81.3& *86.6& *93.7& *84.8& *111.5& *104.0&*131.0 \\
	 0.05 & 41.8& 42.1& *74.9& *101.2& *89.9& *79.4& *103.9& *103.7&*125.7 \\
	 0.1 & 41.0& 41.5& *68.2&*106.4 & *84.9& *72.3& *96.1& *102.7& *119.4\\
	 0.2 & 39.4& 41.1& *56.2& *109.4& *76.5& *63.0& *82.2& *98.8&*109.1 \\
	 0.3 & 37.9& 40.9& *46.3& *105.7& *64.1& *60.8& *63.5& *93.8& *95.1\\
	 0.4 & 36.1& *40.3& *38.4& *100.7& *54.6& *62.3& *51.7& *92.8& *83.8\\
	 0.5 & 34.2& *39.1& 32.4&*94.8 & *42.4& *56.1& *39.9& *85.7& *65.7\\
	 0.6 & 31.6& *36.8& 27.2& *85.6& 34.2& *48.1& 30.7& *71.8& *47.6\\
	 0.7 & 28.4& *32.5& 22.6&*69.8 & 27.6& *32.1& 25.0& *41.9& *32.5\\
	 0.8 & 23.6& 26.0& 17.3& *38.2& 21.1& 20.8& 18.4& 19.3& *23.5\\
	 0.9 & 15.3& 16.0& 10.4& *12.1& 12.6& 11.6& 10.7& 10.0& 13.9\\
	 1 & 0.9& 1.1& 0.8& 0.8& 0.9& 1.0& 0.9& 1.0& 1.0\\
	\hline & \multicolumn{9}{|c|}{Adaptive Timing with Actuation Variation} \\
		\cline{2-10}	 &A&AA&A&AA&A&AA&&& \\
	\hline 0 &*39.5 & 30.0& 37.3&*32.5 &45.3 & *33.5& & & \\
	 0.01 & *39.3& 29.8& 37.1&32.7 & 44.9& *33.7& & & \\
	 0.05 & *38.7& 29.2& 36.1& *31.2& 44.1& *32.4& & & \\
	 0.1 & *37.7& 28.4& 35.1& 29.5& 42.6&30.9 && & \\
	 0.2 & *35.6& 26.8& 33.1& 27.2& 40.0& 28.1& & & \\
	 0.3 & *33.7& 25.2& 31.2& 24.9& 37.4& 25.9& & & \\
	 0.4 & 31.5& 23.6& 29.2& 22.9& 34.8& 23.5& & & \\
	 0.5 & 29.4& 21.8& 27.0& 20.5& 31.8& 21.0& & & \\
	 0.6 & 26.9& 19.7& 24.3& 17.9& 28.6& 18.2& & & \\
	 0.7 & 23.7& 17.2& 21.1& 14.9& 24.5& 15.1& & & \\
	 0.8 & 19.6& 14.0& 16.7& 11.3& 19.2& 11.2& & & \\
	 0.9 & 13.0& 9.2& 10.4&6.9 & 11.8& 6.6& & & \\
	 1 & 0.9&0.9 &0.8& 0.8& 0.9& 0.9& & & \\

\hline
 \end{tabular}
 \vspace{2mm}
\caption[Data table showing the aggregated average results for delay in seconds rounded to the nearest tenth second for select CAV percentages.]{Data table showing the aggregated average results for delay in seconds rounded to the nearest tenth second for select CAV percentages. `R' is the restrictive policy, `p' is the permissive policy, and `c' is the current/real policy. Policies are listed as ``CAV Policy, HV Policy''. `A' represents adaptive timing while ``aa'' represents adaptive timing with actuation. `*' represents that data from AV percentages at this point are a lower bound as the simulator could not spawn vehicles at scheduled times in 1 or more simulation steps due to queue spillbacks.}
 \label{tab:data}
 \end{center}
\end{table*}

\bibliographystyle{IEEEtran}
\bibliography{sigproc} \label{bibliography}

\end{document}

%% file: header.tex
\usepackage{amsmath,amsfonts}
\usepackage{array}
\usepackage{textcomp}
\usepackage{stfloats}
\usepackage{url}
\usepackage{graphicx}
\usepackage{cite}
\usepackage{algorithmic}
\usepackage{algorithm}
\usepackage{amsthm}
\usepackage{amssymb}
\usepackage{array}
\usepackage{bm}
\usepackage{enumerate}
\usepackage{fancyhdr}
\usepackage{fixltx2e}
\usepackage{graphicx}
\usepackage{multirow}
\usepackage{nag}
\usepackage{natbib}
\usepackage{tikz}
\usepackage{ifthen}
\usetikzlibrary{arrows}
\usepackage{url}
\usepackage{verbatim}

\definecolor{huntergreen}{rgb}{0.21, 0.37, 0.23}

%% file: HAIM-technical.bbl
\begin{thebibliography}{}

\bibitem[\protect\citeauthoryear{Au, Shahidi, and Stone}{Au
  et~al.}{2011a}]{liveness}
Au, T.-C., N.~Shahidi, and P.~Stone (2011a).
\newblock Enforcing liveness in autonomous traffic management.
\newblock In {\em Proceedings of the Twenty-Fifth AAAI Conference on Artificial
  Intelligence}, AAAI'11, pp.\  1317–1322. AAAI Press.

\bibitem[\protect\citeauthoryear{Au, Shahidi, and Stone}{Au
  et~al.}{2011b}]{au2011enforcing}
Au, T.-C., N.~Shahidi, and P.~Stone (2011b).
\newblock Enforcing liveness in autonomous traffic management.
\newblock In {\em AAAI}.

\bibitem[\protect\citeauthoryear{Au, Zhang, and Stone}{Au
  et~al.}{2015}]{au2015autonomous}
Au, T.-C., S.~Zhang, and P.~Stone (2015).
\newblock Autonomous intersection management for semi-autonomous vehicles.
\newblock {\em The Routledge Handbook of Transportation\/}, 88--104.

\bibitem[\protect\citeauthoryear{Bansal and Kockelman}{Bansal and
  Kockelman}{2016}]{bansal2016forecasting}
Bansal, P. and K.~M. Kockelman (2016).
\newblock Forecasting americans' long-term adoption of connected and autonomous
  vehicle technologies.
\newblock In {\em Transportation Research Board 95th Annual Meeting}, Number
  16-1871.

\bibitem[\protect\citeauthoryear{Bashiri, Jafarzadeh, and Fleming}{Bashiri
  et~al.}{2018}]{bashiri2018paim}
Bashiri, M., H.~Jafarzadeh, and C.~H. Fleming (2018).
\newblock Paim: Platoon-based autonomous intersection management.
\newblock In {\em 2018 21st International Conference on Intelligent
  Transportation Systems (ITSC)}, pp.\  374--380. IEEE.

\bibitem[\protect\citeauthoryear{Bautista, Dy, Ma{\~n}alac, Orbe, and
  Cordel}{Bautista et~al.}{2016}]{bautista2016convolutional}
Bautista, C.~M., C.~A. Dy, M.~I. Ma{\~n}alac, R.~A. Orbe, and M.~Cordel (2016).
\newblock Convolutional neural network for vehicle detection in low resolution
  traffic videos.
\newblock In {\em 2016 IEEE Region 10 Symposium (TENSYMP)}, pp.\  277--281.
  IEEE.

\bibitem[\protect\citeauthoryear{Bento, Parafita, Santos, and Nunes}{Bento
  et~al.}{2013}]{bento2013intelligent}
Bento, L.~C., R.~Parafita, S.~Santos, and U.~Nunes (2013).
\newblock Intelligent traffic management at intersections: Legacy mode for
  vehicles not equipped with v2v and v2i communications.
\newblock In {\em Intelligent Transportation Systems-(ITSC), 2013 16th
  International IEEE Conference on}, pp.\  726--731. IEEE.

\bibitem[\protect\citeauthoryear{Carlino, Boyles, and Stone}{Carlino
  et~al.}{2013}]{carlino2013auction}
Carlino, D., S.~D. Boyles, and P.~Stone (2013).
\newblock Auction-based autonomous intersection management.
\newblock In {\em Intelligent Transportation Systems-(ITSC), 2013 16th
  International IEEE Conference on}, pp.\  529--534. IEEE.

\bibitem[\protect\citeauthoryear{Chen and Englund}{Chen and
  Englund}{2015}]{chen2015cooperative}
Chen, L. and C.~Englund (2015).
\newblock Cooperative intersection management: A survey.
\newblock {\em IEEE Transactions on Intelligent Transportation Systems\/}~{\em
  17\/}(2), 570--586.

\bibitem[\protect\citeauthoryear{Coifman, Beymer, McLauchlan, and
  Malik}{Coifman et~al.}{1998}]{coifman1998real}
Coifman, B., D.~Beymer, P.~McLauchlan, and J.~Malik (1998).
\newblock A real-time computer vision system for vehicle tracking and traffic
  surveillance.
\newblock {\em Transportation Research Part C: Emerging Technologies\/}~{\em
  6\/}(4), 271--288.

\bibitem[\protect\citeauthoryear{Dresner and Stone}{Dresner and
  Stone}{2008}]{dresner2008multiagent}
Dresner, K. and P.~Stone (2008).
\newblock A multiagent approach to autonomous intersection management.
\newblock {\em Journal of artificial intelligence research\/}~{\em 31},
  591--656.

\bibitem[\protect\citeauthoryear{Fajardo, Au, Waller, Stone, and Yang}{Fajardo
  et~al.}{2011}]{fajardo2011automated}
Fajardo, D., T.-C. Au, S.~Waller, P.~Stone, and D.~Yang (2011).
\newblock Automated intersection control: Performance of future innovation
  versus current traffic signal control.
\newblock {\em Transportation Research Record: Journal of the Transportation
  Research Board\/}~(2259), 223--232.

\bibitem[\protect\citeauthoryear{Feng, Head, Khoshmagham, and Zamanipour}{Feng
  et~al.}{2015}]{feng2015real}
Feng, Y., K.~L. Head, S.~Khoshmagham, and M.~Zamanipour (2015).
\newblock A real-time adaptive signal control in a connected vehicle
  environment.
\newblock {\em Transportation Research Part C: Emerging Technologies\/}~{\em
  55}, 460--473.

\bibitem[\protect\citeauthoryear{FHWA}{FHWA}{2008}]{trafficsignaltimingmanualchap5}
FHWA (2008).
\newblock Traffic signal timing manual.
\newblock Accessed July 23, 2020.\\Available:
  https://ops.fhwa.dot.gov/publications\\/fhwahop08024/chapter5.htm.

\bibitem[\protect\citeauthoryear{FHWA}{FHWA}{2013}]{safety}
FHWA (2013, July).
\newblock Signalized intersections: An informational guide.
\newblock Accessed July 23, 2020. \\Available:
  https://safety.fhwa.dot.gov/intersection\\/conventional/signalized/fhwasa13027/ch5.cfm.

\bibitem[\protect\citeauthoryear{Gajda, Sroka, Stencel, Wajda, and
  Zeglen}{Gajda et~al.}{2001}]{gajda2001vehicle}
Gajda, J., R.~Sroka, M.~Stencel, A.~Wajda, and T.~Zeglen (2001).
\newblock A vehicle classification based on inductive loop detectors.
\newblock In {\em Instrumentation and Measurement Technology Conference, 2001.
  IMTC 2001. Proceedings of the 18th IEEE}, Volume~1, pp.\  460--464. IEEE.

\bibitem[\protect\citeauthoryear{Google}{Google}{2020}]{maps}
Google (2020).
\newblock Google maps.
\newblock Accessed 2020. Available: http://maps.google.com/.

\bibitem[\protect\citeauthoryear{Guler, Menendez, and Meier}{Guler
  et~al.}{2014}]{guler2014using}
Guler, S.~I., M.~Menendez, and L.~Meier (2014).
\newblock Using connected vehicle technology to improve the efficiency of
  intersections.
\newblock {\em Transportation Research Part C: Emerging Technologies\/}~{\em
  46}, 121--131.

\bibitem[\protect\citeauthoryear{Hasch, Topak, Schnabel, Zwick, Weigel, and
  Waldschmidt}{Hasch et~al.}{2012}]{hasch2012millimeter}
Hasch, J., E.~Topak, R.~Schnabel, T.~Zwick, R.~Weigel, and C.~Waldschmidt
  (2012).
\newblock Millimeter-wave technology for automotive radar sensors in the 77 ghz
  frequency band.
\newblock {\em IEEE Transactions on Microwave Theory and Techniques\/}~{\em
  60\/}(3), 845--860.

\bibitem[\protect\citeauthoryear{He, Head, and Ding}{He
  et~al.}{2012}]{he2012pamscod}
He, Q., K.~L. Head, and J.~Ding (2012).
\newblock Pamscod: Platoon-based arterial multi-modal signal control with
  online data.
\newblock {\em Transportation Research Part C: Emerging Technologies\/}~{\em
  20\/}(1), 164--184.

\bibitem[\protect\citeauthoryear{Iwasaki, Kawata, and Nakamiya}{Iwasaki
  et~al.}{2011}]{iwasaki2011robust}
Iwasaki, Y., S.~Kawata, and T.~Nakamiya (2011).
\newblock Robust vehicle detection even in poor visibility conditions using
  infrared thermal images and its application to road traffic flow monitoring.
\newblock {\em Measurement Science and Technology\/}~{\em 22\/}(8), 085501.

\bibitem[\protect\citeauthoryear{Jin, Wu, Boriboonsomsin, and Barth}{Jin
  et~al.}{2013}]{jin2013platoon}
Jin, Q., G.~Wu, K.~Boriboonsomsin, and M.~Barth (2013).
\newblock Platoon-based multi-agent intersection management for connected
  vehicle.
\newblock In {\em 16th International IEEE Conference on Intelligent
  Transportation Systems (ITSC 2013)}, pp.\  1462--1467. IEEE.

\bibitem[\protect\citeauthoryear{Kim and Kumar}{Kim and
  Kumar}{2014}]{kim2014mpc}
Kim, K.-D. and P.~R. Kumar (2014).
\newblock An mpc-based approach to provable system-wide safety and liveness of
  autonomous ground traffic.
\newblock {\em IEEE Transactions on Automatic Control\/}~{\em 59\/}(12),
  3341--3356.

\bibitem[\protect\citeauthoryear{Lee, Park, and Yun}{Lee
  et~al.}{2013}]{lee2013cumulative}
Lee, J., B.~Park, and I.~Yun (2013).
\newblock Cumulative travel-time responsive real-time intersection control
  algorithm in the connected vehicle environment.
\newblock {\em Journal of Transportation Engineering\/}~{\em 139\/}(10),
  1020--1029.

\bibitem[\protect\citeauthoryear{Levin and Boyles}{Levin and
  Boyles}{2015}]{levin2015intersection}
Levin, M.~W. and S.~D. Boyles (2015).
\newblock Intersection auctions and reservation-based control in dynamic
  traffic assignment.
\newblock {\em Transportation Research Record\/}~{\em 2497\/}(1), 35--44.

\bibitem[\protect\citeauthoryear{Liu, Hsieh, and Kumar}{Liu
  et~al.}{2018}]{liu2018safe}
Liu, X., P.-C. Hsieh, and P.~Kumar (2018).
\newblock Safe intersection management for mixed transportation systems with
  human-driven and autonomous vehicles.
\newblock In {\em 2018 56th Annual Allerton Conference on Communication,
  Control, and Computing (Allerton)}, pp.\  834--841. IEEE.

\bibitem[\protect\citeauthoryear{Malikopoulos, Beaver, and
  Chremos}{Malikopoulos et~al.}{2020}]{malikopoulos2020optimal}
Malikopoulos, A.~A., L.~E. Beaver, and I.~V. Chremos (2020).
\newblock Optimal path planning and coordination for connected and automated
  vehicles.

\bibitem[\protect\citeauthoryear{Meyer, Hinz, Laika, Weihing, and Bamler}{Meyer
  et~al.}{2006}]{meyer2006performance}
Meyer, F., S.~Hinz, A.~Laika, D.~Weihing, and R.~Bamler (2006).
\newblock Performance analysis of the terrasar-x traffic monitoring concept.
\newblock {\em ISPRS Journal of Photogrammetry and Remote Sensing\/}~{\em
  61\/}(3-4), 225--242.

\bibitem[\protect\citeauthoryear{Mirzaei and Givargis}{Mirzaei and
  Givargis}{2017}]{mirzaei2017fine}
Mirzaei, H. and T.~Givargis (2017).
\newblock Fine-grained acceleration control for autonomous intersection
  management using deep reinforcement learning.
\newblock In {\em 2017 IEEE SmartWorld, Ubiquitous Intelligence \& Computing,
  Advanced \& Trusted Computed, Scalable Computing \& Communications, Cloud \&
  Big Data Computing, Internet of People and Smart City Innovation
  (SmartWorld/SCALCOM/UIC/ATC/CBDCom/IOP/SCI)}, pp.\  1--8. IEEE.

\bibitem[\protect\citeauthoryear{Mladenovi{\'c} and Abbas}{Mladenovi{\'c} and
  Abbas}{2013}]{mladenovic2013self}
Mladenovi{\'c}, M.~N. and M.~M. Abbas (2013).
\newblock Self-organizing control framework for driverless vehicles.
\newblock In {\em 16th International IEEE Conference on Intelligent
  Transportation Systems (ITSC 2013)}, pp.\  2076--2081. IEEE.

\bibitem[\protect\citeauthoryear{Murgovski, de~Campos, and
  Sj{\"o}berg}{Murgovski et~al.}{2015}]{murgovski2015convex}
Murgovski, N., G.~R. de~Campos, and J.~Sj{\"o}berg (2015).
\newblock Convex modeling of conflict resolution at traffic intersections.
\newblock In {\em 2015 54th IEEE Conference on Decision and Control (CDC)},
  pp.\  4708--4713. IEEE.

\bibitem[\protect\citeauthoryear{Namazi, Li, and Lu}{Namazi
  et~al.}{2019}]{namazi2019intelligent}
Namazi, E., J.~Li, and C.~Lu (2019).
\newblock Intelligent intersection management systems considering autonomous
  vehicles: a systematic literature review.
\newblock {\em IEEE Access\/}~{\em 7}, 91946--91965.

\bibitem[\protect\citeauthoryear{Parks-Young and Sharon}{Parks-Young and
  Sharon}{2022}]{HAIM_Parks-Young_Sharon}
Parks-Young, A. and G.~Sharon (2022).
\newblock Intersection management protocol for mixed autonomous and
  human-operated vehicles.
\newblock {\em IEEE Transactions on Intelligent Transportation Systems\/},
  1--11.

\bibitem[\protect\citeauthoryear{Riegger, Carlander, Lidander, Murgovski, and
  Sj{\"o}berg}{Riegger et~al.}{2016}]{riegger2016centralized}
Riegger, L., M.~Carlander, N.~Lidander, N.~Murgovski, and J.~Sj{\"o}berg
  (2016).
\newblock Centralized mpc for autonomous intersection crossing.
\newblock In {\em 2016 IEEE 19th International Conference on Intelligent
  Transportation Systems (ITSC)}, pp.\  1372--1377. IEEE.

\bibitem[\protect\citeauthoryear{Samczynski, Kulpa, Malanowski, Krysik,
  et~al.}{Samczynski et~al.}{2011}]{samczynski2011concept}
Samczynski, P., K.~Kulpa, M.~Malanowski, P.~Krysik, et~al. (2011).
\newblock A concept of gsm-based passive radar for vehicle traffic monitoring.
\newblock In {\em 2011 MICROWAVES, RADAR AND REMOTE SENSING SYMPOSIUM}, pp.\
  271--274. IEEE.

\bibitem[\protect\citeauthoryear{Sayin, Lin, Shiraishi, Shen, and
  Ba{\c{s}}ar}{Sayin et~al.}{2018}]{sayin2018information}
Sayin, M.~O., C.-W. Lin, S.~Shiraishi, J.~Shen, and T.~Ba{\c{s}}ar (2018).
\newblock Information-driven autonomous intersection control via incentive
  compatible mechanisms.
\newblock {\em IEEE Transactions on Intelligent Transportation Systems\/}~{\em
  20\/}(3), 912--924.

\bibitem[\protect\citeauthoryear{Sharon and Stone}{Sharon and
  Stone}{2017}]{sharon2017protocol}
Sharon, G. and P.~Stone (2017).
\newblock A protocol for mixed autonomous and human-operated vehicles at
  intersections.
\newblock In {\em International Conference on Autonomous Agents and Multiagent
  Systems}, pp.\  151--167. Springer.

\bibitem[\protect\citeauthoryear{UDOT}{UDOT}{2020a}]{udotatspm}
UDOT (2020a).
\newblock Udot automated traffic signal performance measures - automated
  traffic signal performance metrics.
\newblock Accessed July 23, 2020. Available:
  https://udottraffic.utah.gov/ATSPM.

\bibitem[\protect\citeauthoryear{UDOT}{UDOT}{2020b}]{udotopendata}
UDOT (2020b).
\newblock Udot data portal.
\newblock Accessed July 23, 2020. Available:
  http://data-uplan.opendata.arcgis.com/.

\bibitem[\protect\citeauthoryear{VanMiddlesworth, Dresner, and
  Stone}{VanMiddlesworth et~al.}{2008}]{vanmiddlesworth2008replacing}
VanMiddlesworth, M., K.~Dresner, and P.~Stone (2008).
\newblock Replacing the stop sign: Unmanaged intersection control for
  autonomous vehicles.
\newblock In {\em Proceedings of the 7th international joint conference on
  Autonomous agents and multiagent systems-Volume 3}, pp.\  1413--1416.
  International Foundation for Autonomous Agents and Multiagent Systems.

\bibitem[\protect\citeauthoryear{Vasirani and Ossowski}{Vasirani and
  Ossowski}{2012}]{vasirani2012market}
Vasirani, M. and S.~Ossowski (2012).
\newblock A market-inspired approach for intersection management in urban road
  traffic networks.
\newblock {\em Journal of Artificial Intelligence Research\/}~{\em 43},
  621--659.

\bibitem[\protect\citeauthoryear{Wei, Mashayekhy, and Papineau}{Wei
  et~al.}{2018}]{wei2018intersection}
Wei, H., L.~Mashayekhy, and J.~Papineau (2018).
\newblock Intersection management for connected autonomous vehicles: A game
  theoretic framework.
\newblock In {\em 2018 21st International Conference on Intelligent
  Transportation Systems (ITSC)}, pp.\  583--588. IEEE.

\bibitem[\protect\citeauthoryear{Wu, Chen, and Zhu}{Wu
  et~al.}{2019}]{wu2019dcl}
Wu, Y., H.~Chen, and F.~Zhu (2019).
\newblock Dcl-aim: Decentralized coordination learning of autonomous
  intersection management for connected and automated vehicles.
\newblock {\em Transportation Research Part C: Emerging Technologies\/}~{\em
  103}, 246--260.

\bibitem[\protect\citeauthoryear{Yang, Guler, and Menendez}{Yang
  et~al.}{2016}]{yang2016isolated}
Yang, K., S.~I. Guler, and M.~Menendez (2016).
\newblock Isolated intersection control for various levels of vehicle
  technology: Conventional, connected, and automated vehicles.
\newblock {\em Transportation Research Part C: Emerging Technologies\/}~{\em
  72}, 109--129.

\end{thebibliography}
